\documentstyle[12pt]{article}
\newlength{\pubnumber} 

\catcode`\@=11
\@addtoreset{equation}{section}
\def\section{\@startsection{section}{1}{\z@}{3.5ex plus 1ex minus .2ex}
 {2.3ex plus .2ex}{\large\bf}}
\def\subsection{\@startsection{subsection}{2}{\z@}{2.3ex plus .2ex}
 {2.3ex plus .2ex}{\bf}}

\begin{document}

\begin{titlepage}
\samepage{
\setcounter{page}{1}
\rightline{IASSNS-HEP-95/53}
\rightline{\tt hep-th/9506388}
\rightline{June 1995}
\vfill
\begin{center}
 {\Large \bf Top Quark Mass Prediction in \\
     Superstring Derived Standard--like Models\\}
\vfill
 {\large Alon E. Faraggi\footnote{
   E-mail address: faraggi@sns.ias.edu}\\}
\vspace{.12in}
 {\it  School of Natural Sciences, Institute for Advanced Study\\
  Olden Lane, Princeton, N.J.~~08540~ USA\\}
\end{center}
\vfill
\begin{abstract}
  {\rm
A remarkable achievement of the realistic superstring
standard--like models is the successful prediction of the top quark mass,
assuming the Minimal Supersymmetric Standard Model spectrum below
the string scale.
Recently it was shown that string scale
unification requires the existence of additional matter, in vector--like
representations, at intermediate energy scales and that certain string
models contain the needed representations in their massless spectrum.
I obtain the top, bottom and tau lepton Yukawa couplings in these models,
by calculating tree level string amplitudes, in terms of the
unified gauge coupling and certain
vacuum expectation values that are required for the consistency of the
string models.
Using two--loop renormalization group equations for the gauge and Yukawa
couplings, I study the effect of the intermediate matter thresholds
on the top quark mass prediction.
Agreement with the experimental values of $\alpha_{\rm strong}(M_Z)$,
$\sin^2\theta_W(M_Z)$ and $\alpha^{-1}_{\rm em}(M_Z)$ is imposed.
It is found that the physical top quark mass prediction
is increased to the range
$192-200$ GeV and that the ratio $\lambda_b(M_Z)/\lambda_\tau(M_Z)$
is in agreement with experiment.
}
\end{abstract}
\vfill
\smallskip}
\end{titlepage}

\setcounter{footnote}{0}

\def\beq{\begin{equation}}
\def\eeq{\end{equation}}
\def\beqn{\begin{eqnarray}}
\def\eeqn{\end{eqnarray}}
\def\AEF{A.E. Faraggi}
\def\NPB#1#2#3{{\it Nucl.\ Phys.}\/ {\bf B#1} (19#2) #3}
\def\PLB#1#2#3{{\it Phys.\ Lett.}\/ {\bf B#1} (19#2) #3}
\def\PRD#1#2#3{{\it Phys.\ Rev.}\/ {\bf D#1} (19#2) #3}
\def\PRL#1#2#3{{\it Phys.\ Rev.\ Lett.}\/ {\bf #1} (19#2) #3}
\def\PRT#1#2#3{{\it Phys.\ Rep.}\/ {\bf#1} (19#2) #3}
\def\MODA#1#2#3{{\it Mod.\ Phys.\ Lett.}\/ {\bf A#1} (19#2) #3}
\def\IJMP#1#2#3{{\it Int.\ J.\ Mod.\ Phys.}\/ {\bf A#1} (19#2) #3}
\def\nuvc#1#2#3{{\it Nuovo Cimento}\/ {\bf #1A} (#2) #3}
\def\etal{{\it et al,\/}\ }
\hyphenation{su-per-sym-met-ric non-su-per-sym-met-ric}
\hyphenation{space-time-super-sym-met-ric}
\hyphenation{mod-u-lar mod-u-lar--in-var-i-ant}


\setcounter{footnote}{0}

One of the intriguing successes of the realistic superstring standard--like
models in the free fermionic formulation is the successful prediction
of the top quark mass. In Ref.~\cite{TOP} the top quark mass was predicted
to be in the mass range
\begin{equation}
m_t\approx175-180~{\rm GeV}~,
\label{topmass1991prediction}
\end{equation}
three years prior to its experimental observation.
Remarkably, this prediction is in agreement with the top quark mass
as observed by the recent CDF and D0 collaborations \cite{CDF}.
In obtaining the top quark mass prediction, it was assumed in Ref.
\cite{TOP} that the spectrum below the string scale is that of the Minimal
Supersymmetric Standard Model (MSSM), i.e. three generations plus
two Higgs doublets. However, it was recently shown \cite{DF} that string
gauge coupling unification requires the existence of intermediate
matter thresholds, beyond the MSSM spectrum. This additional matter
takes the form of additional color triplets and electroweak doublets,
in vector--like representations, with specific weak hypercharge
assignments. Remarkably, the same string standard--like models
that led to the prediction, Eq.\ (1), allow for the existence
of the needed additional states and the required weak hypercharge
assignments, to achieve string scale unification.

In this paper, I investigate the effect of the intermediate
matter thresholds on the top quark mass prediction in the realistic
superstring derived standard--like models.
In these models the top, bottom and tau lepton Yukawa couplings
are calculated by evaluating the string tree level amplitudes
between the vertex operators in the effective conformal
field theory. The Yukawa couplings are obtained in terms of the
unified gauge coupling and certain VEVs that are required for the
consistency of the string models. The gauge and Yukawa couplings
are extrapolated numerically from the string unification scale to low
energies by using the coupled two--loop
supersymmetric Renormalization Group Equations (RGEs), including
the contribution of the extra matter thresholds.
Agreement with $\alpha_{\rm strong}(M_Z)$, $\sin^2\theta_W(M_Z)$ and
$\alpha_{\rm em}(M_Z)$ is imposed. It is found that the
running top quark mass is shifted to $m_t\approx185-190$ GeV
and the physical top quark mass is in the range $192-200$ GeV.
The string models under consideration also predict
$\lambda_b=\lambda_\tau$ at the string unification scale.
Using the same extrapolation, it is found that for a large portion of
the parameter space this prediction is in agreement with the
experimental value of $m_b(M_Z)/m_\tau(M_Z)$.
Thus, I show that LEP precision data for
$\alpha_{\rm strong}$ and $\sin^2\theta_W$ as well as the CDF/D0 top
quark observation and the $b/\tau$ mass relation can all simultaneously be
consistent with the superstring derived standard--like models.
Here I present the main results. A detailed account of
the numerical results and details of the string calculations of the Yukawa
couplings will be given elsewhere \cite{bigpaper}.

One of the fundamental mysteries of the observed fermion mass spectrum is
the large mass splitting between the top quark and the lighter
quarks and leptons. Especially difficult
to understand within the context of the Standard Model,
and its field theoretic
extensions, is the big splitting in the heaviest generation.
The superstring derived standard--like models
suggest a superstring mechanism that explains the suppression of the
lighter quarks and leptons masses relative to the top quark mass.
At the cubic level of the superpotential only $+{2\over3}$ charged
quarks get nonvanishing Yukawa couplings, while the remaining
quarks and leptons get their mass terms from nonrenormalizable
terms. This selection mechanism, between $+{2\over3}$ and
$-{1\over3}$ Yukawa couplings, results from the specific assignments
of boundary conditions that specify the string models.
Due to the horizontal symmetries of the string models,
each of the chiral generations couples at the cubic level to different
doublet Higgs multiplets.
Only one pair of the Higgs doublets remains light at low energies \cite{FM}.
As a result only one nonvanishing mass term, namely
the top quark mass term, remains at low energies. The mass terms
for the lighter quarks and leptons are obtained from nonrenormalizable terms.
The nonrenormalizable terms have the general
form,
\begin{equation}
cgf_if_jh(\phi/M)^{n-3}
\label{generalnonrenoterm}
\end{equation}
where $c$ are the calculable coefficients of the $n^{\rm th}$
order correlators, $g$ is the gauge coupling at the unification scale,
$f_i$, $f_j$ are the quark and lepton fields, $h$ are the light
Higgs representations, and $\phi$ are Standard Model singlets in the
massless spectrum of the string models.
An important property of the superstring standard--like models is the
absence of gauge and gravitational anomalies apart from a single ``anomalous
$U(1)$" symmetry. This anomalous $U(1)_A$ generates a Fayet--Iliopoulos term
that breaks supersymmetry at the Planck scale \cite{DSW}.
Supersymmetry is restored
and $U(1)_A$ is broken by giving VEVs to a set of
Standard Model singlets in the
massless string spectrum along the flat F and D directions \cite{DSW}.
However, as the charge of these singlets must have $Q_A<0$ to cancel
the anomalous
$U(1)$ D--term equation, in  many models a phenomenologically realistic
solution does not exist. The only models that
were found to admit a solution are models which have cubic level
Yukawa couplings only for $+{2\over3}$ charged quarks.
The magnitude of these VEVs is set by the Fayet--Iliopoulos D--term, which
is generated due to the ``anomalous'' $U(1)_A$ at the one--loop level in
string perturbation theory.
Consequently, some of the nonrenormalizable, order $N$ terms,
become effective renormalizable terms
with effective Yukawa couplings, $\lambda=cg(\langle\phi\rangle/M)^{n-3}$.

The superstring standard--like
models are constructed in the free fermionic formulation \cite{freefermions}.
A model is generated by a consistent set of boundary
condition basis vectors.
The physical spectrum is obtained
by applying the generalized GSO projections. Each physical state
is described in terms of vertex operators in the effective
conformal field theory. Quark and lepton mass terms are obtained
by calculating the correlators
\begin{equation}
A_N\sim\langle V_1^fV_2^fV_3^b\cdot\cdot\cdot V_N^b\rangle,
\label{orderncorrelators}
\end{equation}
between the vertex operators.
The nonvanishing correlators must be invariant under all the symmetries
of a given string model and satisfy all the string selection rules.
Consequently, most of the quark and lepton mass terms vanish at
the cubic level of the superpotential. One must then examine whether
potential quark and lepton mass terms can be obtained from
nonrenormalizable terms.

The first five basis vectors consist of the NAHE set,
$\{{\bf 1}, S,b_1,b_2,b_3\}$ \cite{nahe}.
At the level of the NAHE set
the gauge group is $SO(10)\times SO(6)^3\times E_8$, with 48 generations.
The number of generations is reduced to three and the $SO(10)$
gauge group is broken to
$SU(3)\times SU(2)\times U(1)^2$
by adding
to the NAHE set three additional basis vectors, $\{\alpha,\beta,\gamma\}$.
The basis vector that breaks the $SO(2n)$ symmetry to $SU(n)\times U(1)$
must contain half integral boundary conditions for the world--sheet
complex fermions that generate the $SO(10)$ symmetry.
This basis vector plays an important role in the superstring
selection mechanism and will be denoted as the vector $\gamma$.

The two basis vectors $\{{\bf 1},S\}$ produce a model with $N=4$
space--time supersymmetry and $SO(44)$ gauge group.
At this level all of the world--sheet fermions are equivalent.
The NAHE set plus the vector $2\gamma$ divide the world--sheet
fermions into several groups. The six left--moving real fermions,
$\chi^{1,\cdots,6}$ are paired to form three complex fermions
denoted $\chi^{12}$, $\chi^{34}$ and $\chi^{56}$.
The sixteen
right--moving complex fermions $\bar\psi^{1\cdots5}\bar\eta^1,
\bar\eta^2,\bar\eta^3,\bar\phi^{1,\cdots,8}$ produce the observable
and hidden gauge groups, that arise from the sixteen dimensional
compactified space of the heterotic string in ten dimensions.
Finally, the twelve left--moving
$\{y,\omega\}^{1\cdots6}$ and
twelve right--moving $\{{\bar y},\bar\omega\}^{1\cdots6}$ real fermions
correspond to the left/right symmetric internal conformal field
theory of the heterotic string.
The assignment of boundary conditions
in the vector $\gamma$ to this set of internal world--sheet fermions
selects cubic level Yukawa couplings for $+{2\over3}$ or $-{1\over3}$
charged quarks.

Each of the sectors $b_1$, $b_2$ and $b_3$ produce one generation.
Three pairs of electroweak Higgs doublets
$\{h_1, h_2, h_3, {\bar h}_1, {\bar h}_2, {\bar h}_3\}$ are obtained
from the Neveu--Schwarz sector. One or two additional pairs, $\{h_{45},
\bar h_{45}\}$ are
obtained from a combination of the additional basis vectors.
The three boundary condition basis vectors $\{\alpha,\beta,\gamma\}$
break the horizontal $SO(6)^3$ symmetries to factors of $U(1)$s.
Three $U(1)$ symmetries arise from the complex right--moving
fermions $\bar\eta^1$, $\bar\eta^2$, $\bar\eta^3$. Additional horizontal
$U(1)$ symmetries arise by pairing two of the right--moving real
internal fermions $\{{\bar y},{\bar\omega}\}$. For
every right--moving $U(1)$ symmetry, there is a corresponding
left--moving global $U(1)$ symmetry that is obtained by pairing two of the
left--moving real fermions $\{y,\omega\}$. Each of the remaining
world--sheet internal fermions from the set $\{y,\omega\}$ is paired
with a right--moving real internal fermions from the set
$\{{\bar y},{\bar\omega}\}$ to form a Ising model operator.

The assignment of boundary conditions in the basis vector $\gamma$
for the internal world--sheet fermions,
$\{y,\omega\vert{\bar y},{\bar\omega}\}$ selects a cubic level mass term
for $+{2\over3}$ or $-{1\over3}$ charged quarks. For each of the
sectors $b_1$, $b_2$ and $b_3$ the fermionic boundary conditions
select the cubic level Yukawa couplings according to the difference,
\begin{equation}
\Delta_j=\vert\gamma(L_j)-\gamma(R_j)\vert=0,1
\label{updownselectionrule}
\end{equation}
where $\gamma(L_j)/\gamma(R_j)$ are the boundary conditions in the vector
$\gamma$ for the internal world--sheet fermions from the set
$\{y,\omega\vert{\bar y},{\bar\omega}\}$, that are periodic in the vector
$b_j$. If $\Delta_j=1$ then a Yukawa coupling for the $+{2\over3}$ charged
quark from the sector $b_j$ is nonzero and the Yukawa coupling for the
$-{1\over3}$ charged quark vanishes. The opposite occurs if
$\Delta_j=0$. Thus, the states from each of the sectors $b_1$, $b_2$ and
$b_3$ can have a cubic level Yukawa coupling for the $+{2\over3}$ or
$-{1\over3}$ charged quark, but not for both. We can construct string models
in which both $+{2\over3}$ and $-{1\over3}$ charged quarks get a cubic
level mass term. The model of ref. \cite{FNY} is an example of such a model.
By contrast, we can also construct string models in which only
$+{2\over3}$ charged quarks get a nonvanishing cubic level mass term.
The model of ref. \cite{TOP}
is an example of such a model. In Ref. \cite{YUKAWA} this
selection rule is proven by using the string consistency constraints and
Eq.\ (\ref{updownselectionrule})
to show that either the $+{2\over3}$ or the $-{1\over3}$ mass term is
invariant under the $U(1)_j$ symmetry.

Due to the horizontal $U(1)$ symmetries the states from each of the sectors
$b_j$, ($j=1,2,3$) can couple at the cubic level
only to one of the Higgs pairs
$h_j$, $\bar h_j$. This results due to the fact that the states from a sector
$b_j$ and the Higgs doublets $h_j$ and $\bar h_j$ are charged with respect
to one of the horizontal $U(1)_j$, $j=1,2,3$ symmetries.
Therefore, the basis vectors of Ref. \cite{TOP}
lead to the following nonvanishing
cubic level mass terms for the states from the sectors
$b_1$, $b_2$ and $b_3$,
 \begin{equation}
\{{u_{L_i}^c}Q_i{\bar h}_i+{N_{L_i}^c}L_i{\bar h}_i\}~~~~(i=1,2,3),
\label{cubiclevelmassterms}
\end{equation}
where the common normalization factor is obtained by evaluating the
correlators, Eq.\ (\ref{orderncorrelators}), of the $N=3$ order terms in
Eq.\ (\ref{cubiclevelmassterms}), yielding
$A_3=\sqrt{2}g~.$
A very restricted class of standard--like
models with $\Delta_j=1$ for $j=1,2,3$, were found to admit a solution
to the F and D flatness constraints. Consequently, in these models
only $+{2\over3}$ charged quarks obtain a cubic level mass term.
Mass terms for  $-{1\over3}$ charged quarks and for charged leptons
must arise from nonrenormalizable terms. In the model of Ref. \cite{TOP}
the following nonvanishing mass terms are obtained at the
quartic order,
\begin{eqnarray}
W_4=
\{&{d_{L_1}^c}Q_1h_{45}\Phi_1+
   {e_{L_1}^c}L_1h_{45}\Phi_1+\nonumber\\
  &{d_{L_2}^c}Q_2h_{45}{\bar\Phi}_2
+{e_{L_2}^c}L_2h_{45}{\bar\Phi}_2\}.
\label{quarticordermassterms}
\end{eqnarray}
The quartic term coefficients are obtained by calculating the $N=4$
order correlators and are equal to $gI/\sqrt{\pi}M_{\rm Pl}$. $I$ is
a one dimensional complex integral which is evaluated numerically,
$I\approx77.7$.

 From Eq.\ (\ref{quarticordermassterms})
we observe that if some of the Standard Model singlets, that appear
in the quartic order terms, acquire a VEV by the cancelation of the
anomalous $U(1)$ D--term equation, then effective mass terms for the
$-{1\over3}$ quarks and for charged leptons are obtained. At the same
time an analysis of the renormalizable and nonrenormalizable
superpotential suggests that only one pair of Electroweak Higgs doublets
remains light at low energies \cite{FM}. For typical scenarios, those consist
of ${\bar h}_1$ or ${\bar h}_2$ and $h_{45}$.
Because the states from each of the sectors
$b_1$, $b_2$ and $b_3$ can couple only to one of the Higgs pairs, at the
cubic level of the superpotential only one mass term remains.
Therefore, at the cubic level of the superpotential only the
top quark has a nonvanishing mass term.
The top quark Yukawa coupling is therefore given by
\begin{equation}
\lambda_t(M_{\rm string})=g\sqrt2
\label{topyuakawa}
\end{equation}
where $g$ is the gauge coupling at the unification scale.

In ref. \cite{TOP} a solution to the F and D flatness
constraints was found with,
\begin{equation}
\vert\langle{\bar\Phi}_2\rangle\vert^2=
{{g^2}\over{16\pi^2}}{1\over{ 2\alpha^\prime}}
\end{equation}
where $\alpha^\prime$ is the string tension,
$\alpha^\prime={{16\pi}/{g^2M_{pl}^2}}$ \cite{KLN}.
Thus, after
inserting the VEV of ${\bar\Phi}_2$, the effective bottom quark and tau lepton
Yukawa couplings are given by,
\begin{equation}
\lambda_b=\lambda_\tau=0.35g^3.
\label{btauyukawa}
\end{equation}
The top quark mass prediction is obtained by taking
$g\sim{1/{\sqrt2}}$ at the unification scale.
The three Yukawa couplings are run
to the low energy scale by using the MSSM one--loop RGEs. The bottom quark
mass, $m_b(M_Z)$
and the $W$-boson mass, $M_{W}(M_{ W})$
are used to fix the two VEVs, $v_1$ and $v_2$.
Using the relation,
\begin{equation}
m_t\approx\lambda_t(M_Z)\sqrt{{{2M_W^2}\over{g^2_2(M_W)}}-
\left({{m_b(M_Z)}\over{\lambda_b(m_Z)}}\right)^2}\label{topbottommassrelation}
\end{equation}
the top quark mass prediction,
Eq.\ (\ref{topmass1991prediction}), is obtained.

Eq.\ (\ref{topmass1991prediction}) was obtained in ref. \cite{TOP},
assuming the MSSM spectrum below the string unification scale,
$M_{\rm string}\approx g_{\rm string}\times5\times10^{17}$ GeV, where
$g_{\rm string}$ is the gauge coupling at the string unification scale.
However, this assumption results in disagreement with the values
extracted at LEP for $\alpha_{\rm strong}(M_Z)$ and
$\sin^2\theta_W(M_Z)$.
In ref. \cite{DF} it was shown, in a wide range of realistic
free fermionic models, that heavy string threshold corrections,
non-standard hypercharge normalizations,
light SUSY thresholds or intermediate
gauge structure, do not resolve the disagreement with
$\alpha_{\rm strong}(M_Z)$ and $\sin^2\theta_W(M_Z)$.
The problem may be resolved in the superstring derived standard--like models
due to the existence of color triplets and electroweak doublets
from exotic sectors that arise from the additional vectors $\alpha$,
$\beta$ and $\gamma$.
For example, the model of Ref. \cite{GCU} is obtained from the model
of ref. \cite{TOP} by a change of GSO phase that preserves the observable
massless spectrum and interactions. This model contains in its spectrum
two pairs of $(\overline{3},1)_{1/3}$
color triplets with beta-function coefficients $(b_3,b_2,b_1)=(1/2,0,1/5)$,
one pair of $(\overline{3},1)_{1/6}$ triplets with $b_i=(1/2,0,1/20)$,
and three pairs of $(1,2)_{0}$ doublets with $b_i=(0,1/2,0)$.
This particular combination of representations and hypercharge
assignments opens up a sizable window
in which the low-energy data and string unification can
be reconciled.  For example, it is found that if these triplets
all have equal masses in the approximate range
$ 2\times 10^{11} \leq M_3 \leq 7\times 10^{13}$ GeV
with the doublet masses in the corresponding range
$ 9\times 10^{13} \leq M_2 \leq 7\times 10^{14}$ GeV,
then the discrepancy is removed.

This extra matter at intermediate energy scales may also affect the top quark
mass prediction, Eq.\ (\ref{topmass1991prediction}).
To study this effect, I take the
two--loop RGEs for the gauge and Yukawa couplings, including
the contribution of the
extra matter. To account for the dependence of
$M_{\rm string}$ on $g_{\rm string}$,
$M_{\rm string}$ is varied in the range $(3-7)\times10^{17}$ GeV.
The Yukawa couplings at
$M_{\rm string}$ are given by Eqs. (\ref{topyuakawa},\ref{btauyukawa}), and
$g_{\rm string}$ is varied in the range $0.03-0.07$.
The two--loop RGEs are then evolved to the extra doublets and triplets
thresholds. The extra doublet and triplet thresholds are varied in the ranges
$ 1\times 10^{13} \leq M_2 \leq 1\times 10^{16}$ GeV and
$ 9\times 10^{9} \leq M_3 \leq 1\times 10^{12}$ GeV, respectively.
The contribution of each threshold to the $\beta$--function coefficients
is removed in a step approximation. The two--loop RGEs are then evolved to the
approximate top quark mass scale, $m_t\approx175$ GeV.
At this scale the top quark Yukawa coupling and
$\alpha_{\rm strong}(m_t)$ are extracted,
and the contribution of the top quark to the RGEs is removed.
The two--loop RGEs
are then evolved to the $Z$ mass scale and agreement with
the experimental values of $\alpha_{\rm strong}(M_Z)=0.12\pm0.01$,
$\sin^2\theta_W(M_Z)=0.232\pm0.001$
and $\alpha^{-1}_{\rm em}(M_Z)=127.9\pm0.1$ is imposed.
The bottom quark and tau lepton masses, $m_b(m_b)=4.3\pm0.2$ and
		$m_\tau(m_\tau)=1777.1^{+0.4}_{-0.5}$ MeV \cite{expalpha} are
evolved from their physical mass scale to the $Z$--mass scale by using the
three--loop QCD and two--loop QED RGEs \cite{BBO}.
The bottom mass is then used to extract the
running top quark mass, using Eq.\ (\ref{topbottommassrelation}).
The physical top quark mass is given by,
$m_t(physical)=m_t(m_t)(1+{4\over{3\pi}}\alpha_{\rm strong}(m_t))$
where $m_t(m_t)$ is given by Eq.\ (\ref{topbottommassrelation}).
It is found
that the effect of the intermediate matter
thresholds is to push $m_t(m_t)$ to the mass range
$m_t(m_t)\approx185-190$ GeV.
The physical top quark mass is in the interval
\begin{equation}
m_t(physical)\approx 192-200 {\rm GeV},
\end{equation}
which is in agreement with the CDF and D0 results.

 From Eq.\ (\ref{btauyukawa}) we observe that in this model
$\lambda_b=\lambda_\tau$ at the string unification scale.
Consequently, an additional prediction for the mass ratio
\begin{equation}
\lambda_b(M_Z)/\lambda_\tau(M_Z)=m_b(M_Z)/m_\tau(M_Z)
\end{equation}
is obtained.
$\lambda_b$ and $\lambda_\tau$ are extrapolated from the string unification
scale to the $Z$--mass scale using the two--loop RGEs with the intermediate
matter thresholds, as described above. The gauge couplings of
$SU(3)_{\rm color}\times U(1)_{em}$ are then extrapolated to the bottom
quark mass scale. The bottom quark and tau lepton masses are then extrapolated
to the $Z$ mass scale. It is then found that the predicted ratio of
\begin{equation}
\lambda_b(M_Z)/\lambda_\tau(M_Z)\approx1.86-1.98
\end{equation}
is in good agreement with the
extrapolated value of $m_b(M_Z)/m_\tau(M_Z)\approx1.57-1.93$.

In this paper I have shown that LEP precision data for
$\alpha_{\rm strong}$ and $\sin^2\theta_W$ as well as the CDF/D0 top
quark observation and the $b/\tau$ mass relation can all simultaneously be
consistent with
the superstring derived standard--like models. This is achieved if
the additional matter states that are obtained in the string derived models
exist at the appropriate thresholds. It will be of further interest to examine
whether these extra matter states have additional testable predictions
and whether the remaining fermion mass spectrum
can be derived from the superstring standard--like models.
Such work is in progress
and will be reported in future publications.

\bigskip
\medskip
\leftline{\large\bf Acknowledgments}
\medskip

I thank K. Dienes and J. Pati for discussions.
This work was supported in part by DOE Grant No.\ DE-FG-0290ER40542.


\vfill\eject

\bibliographystyle{unsrt}

\end{document}